\documentclass[twocolumn,showpacs,preprintnumbers,amsmath,amssymb]{revtex4}

\usepackage{graphicx}
\usepackage{rotating}

\begin{document}

\draft

\title{Crossover from a molecular Bose-Einstein condensate to a degenerate Fermi gas}

\author{M. Bartenstein,$^{1}$ A. Altmeyer,$^{1}$ S. Riedl,$^{1}$ S. Jochim,$^{1}$ C. Chin,$^{1}$ J. Hecker Denschlag,$^{1}$ R.
Grimm$^{1,2}$}

\address{$^{1}$Institut f\"ur Experimentalphysik, Universit\"at Innsbruck, Technikerstra{\ss}e 25, 6020 Innsbruck,
Austria\\$^{2}$Institut f\"ur Quantenoptik und Quanteninformation,
\"Osterreichische Akademie der Wissenschaften, 6020 Innsbruck,
Austria}

\date{\today}


\begin{abstract}
We demonstrate a reversible conversion of a $^6$Li$_2$ molecular
Bose-Einstein condensate to a degenerate Fermi gas of atoms by
adiabatically crossing a Feshbach resonance. By optical in situ
imaging, we observe a smooth change of the cloud size in the
crossover regime. On the Feshbach resonance, the ensemble is
strongly interacting and the measured cloud size is $75(7)\%$ of
the one of a non-interacting zero-temperature Fermi gas. The high
condensate fraction of more than $90\%$ and the adiabatic
crossover suggest our Fermi gas to be cold enough to form a
superfluid.
\end{abstract}

\maketitle

\narrowtext

Bose-Einstein condensation (BEC) of molecules formed by fermionic
atoms was recently demonstrated \cite{li2becinn, k2bec, li2becmit,
private}. The tunability of interactions in such systems provides
a unique possibility to explore the Bose-Einstein condensate to
Bardeen-Cooper-Schrieffer (BEC-BCS) crossover \cite{becbcs}, an
intriguing interplay between the superfluidity of bosons and
Cooper pairing of fermions. While the BEC and BCS limits are both
well understood, the crossover takes place in a strongly
interacting regime, which represents a challenge for many-body
theory.

Feshbach resonances \cite{feshbach} play a central role to control
two-body interaction and have been used for conversion between
fermionic atoms and bosonic molecules \cite{jin, hulet,
li2thermal1, li2thermal2}. They are also the experimental key to
investigate phenomena related to the BEC-BCS crossover. For
example, it has been predicted in Ref.\cite{carr} that a pure
molecular BEC can be converted into a superfluid Fermi gas by an
adiabatic passage over the Feshbach resonance. Moreover, in the
crossover regime where the interactions are unitarity limited, a
universal behavior is expected \cite{heiselberg, ho03}. Ultracold
gases in that regime may provide new insights into other
strongly-interacting systems such as high-$T_c$ superconductors,
$^3$He superfluids, and neutron stars.

A spin-mixture of $^6$Li atoms in the lowest two hyperfine
sub-levels is an excellent system to investigate the crossover
\cite{thomasmechan, salomon} based on a broad Feshbach resonance
at a magnetic field of $B=850$G \cite{feshbach8500, feshbach850,
feshbach850b}. An efficient formation of ultracold molecules has
been realized by three-body recombination \cite{li2thermal2,
Chin2003}, or by sweeping the magnetic field across the resonance
\cite{li2thermal1}. The long lifetime of the molecules permits
efficient evaporation \cite{li2thermal1, li2thermal2, li2becinn}
and facilitates slow, adiabatic changes of the system.

In this work, we explore the regime where the BEC-BCS crossover is
expected by analyzing the density profiles of the trapped cloud at
different magnetic fields. Our experimental setup is described in
Ref.~\cite{li2becinn}. We load $2\times 10^6$ precooled $^6$Li
atoms into a single focused-beam dipole trap, which is generated
by a 10W Yb:YAG laser operating at a wavelength of 1030\,nm. We
evaporatively cool the cloud by exponentially lowering the trap
depth with a time constant of $460$ms. The radial and axial trap
frequencies are $\omega_r/2\pi=110$Hz$(P/{\rm mW})^{1/2}$ and
$\omega_z/2\pi=(600 B/{\rm kG}+0.94 P/{\rm mW})^{1/2}$Hz,
respectively, where $P$ is the laser power. The curvature of the
magnetic field that we use for Feshbach tuning results in a
magnetic contribution to the axial trapping. In the low power
range where the molecular BEC is formed ($P < 50$\,mW), the axial
confinement is predominantly magnetic. During the whole
evaporation process the magnetic field is kept at $B=764$G. At
this field the molecular binding energy is $\sim$\,$k_B \times
2\mu$K, where $k_B$ is Boltzmann's constant. For the scattering
length of elastic molecule-molecule collisions we expect $a_{mol}
= 2200\,a_0$, based on the predicted relation of $a_{mol} = 0.6a$
\cite{petrovam} and an atomic scattering length of $a = 3500\,a_0$
\cite{feshbach850}. Here $a_0$ is Bohr's radius. Using
radio-frequency spectroscopy which allows us to distinguish
signals from atoms and molecules \cite{jin}, we observe a complete
atom to molecule conversion when the thermal energy of the
particles is reduced to values well below the molecular binding
energy.

\begin{figure}
\includegraphics[width=3in]{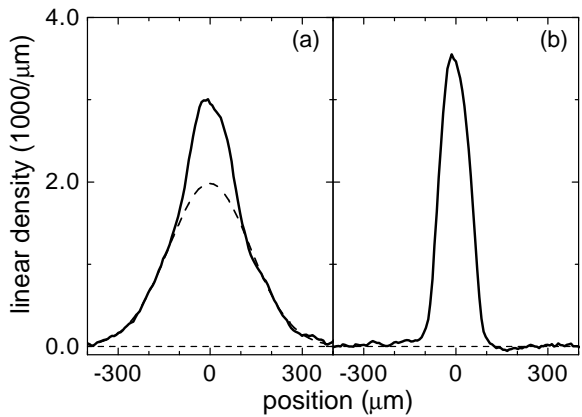}
\caption{Axial density profiles of a partially condensed (a) and
fully condensed (b) molecular cloud. The profiles are derived from
averaging seven in situ images taken at a magnetic field of
$B=676$G after evaporation at the production field of 764G. (a)
When the evaporation ramp is stopped with $4\times 10^5$ molecules
at a final laser power of 28\,mW, a characteristic bimodal
distribution is observed with a condensate fraction of $\sim$20\%.
The dashed curve shows Gaussian fit to the thermal fraction. (b)
At a final laser power of 3.8\,mW, an essentially pure condensate
of $2\times 10^5$ molecules is obtained.} \label{fig1}
\end{figure}

For detection we apply \emph{in situ} absorption imaging to record
spatial density profiles of the trapped ensemble. To image at high
magnetic fields, we illuminate the cloud for $20\mu$s with a probe
beam (intensity $0.5$mW/cm$^2$) tuned to the atomic $|2S_{1/2},
m_J=-1/2, m_I=0\rangle \rightarrow |2P_{3/2}, m'_J=-3/2, m'_I=0
\rangle$ transition. The probe beam dissociates the molecules and
is used to image the resulting atom cloud \cite{li2becmit}.
Compared to the absorption imaging of unbound atoms, we found that
the detection efficiency of the molecules approaches $100\%$ at
fields higher than 750G and $\sim 50\%$ at 650G. The difference is
due to the Frank-Condon wavefunction overlap, which favors fields
closer to the resonance where the interatomic separation in the
molecular state is larger. In our cigar-shaped trap, the radial
cloud size is on the order of our imaging resolution of $10\mu$m,
while the axial cloud size of typically $\sim 100\mu$m can be
accurately measured. We therefore obtain axial density
distributions from images integrated radially.

To measure the condensate fraction, we adiabatically reduce the
magnetic field from 764\,G to 676\,G in a 200-ms linear ramp after
completion of the evaporation ramp. This reduces the scattering
length $a_{mol}$ and thus increases the visibility of the
characteristic bimodal distribution. Fig.~\ref{fig1}(a) shows a
bimodal profile observed in this way with
$N_{mol}=N/2=4\times10^5$ molecules remaining at a final
evaporation ramp power of 28\,mW. A Gaussian fit to the thermal
wings (dashed line) yields a temperature of T=430\,nK, which is a
factor of 7.5 below the calculated trap depth of 3.2\,$\mu$K. The
observed condensate fraction of $\sim20\%$ is consistent with
$1-(T/T_c)^3$, where
$T_c=0.8k_B^{-1}\hbar\bar{\omega}(N_{mol}/1.202)^{1/3}=$500\,nK is
the critical temperature,
$\bar{\omega}=(\omega_r^2\omega_z)^{1/3}$ is the mean vibration
frequency, and the factor of 0.8 takes into account the $\sim$20\%
down-shift in $T_c$ due to interactions \cite{giorgini1996a}.

We obtain pure molecular condensates when we continue the
evaporation process down to final power levels of a few mW.
Fig.~\ref{fig1}(b) shows an essentially pure condensate of
$N_{mol}=2.0\times10^5$ molecules obtained at a final power of
3.8\,mW, where the trap depth is $450$nK. The density profile is
well fit by a Thomas-Fermi density distribution $\propto
(1-z^2/z^2_{\rm TF})^2$ with a radius $z_{\rm TF} = 105\,\mu$m.
The corresponding peak molecular density is $1.2\times
10^{13}$cm$^{-3}$. In the image a thermal component is not
discernable. A careful analysis of the profile provides us with a
lower bound of $90\%$ for the condensate fraction. For the
chemical potential of the BEC we obtain $\mu = \frac{1}{2}m_{\rm
mol}\omega_z^2 z^2_{\rm TF} = k_{\rm B} \times 130$\,nK. Here
$m_{\rm mol} = 2m$ is the mass of the $^6$Li dimer. Based on the
prediction $a_{mol}=0.6a=650a_0$, the calculated chemical
potential of
$\frac12(15\hbar^2N_{mol}\bar{\omega}^3a_{mol}\sqrt{m_{mol}})^{2/5}=k_B\times
155$\,nK is consistent with the observed value of $k_B\times
130$\,nK considering the experimental uncertainty. In particular,
the particle number is calibrated to within a factor of $1.5$
through fluorescence imaging \cite{li2thermal2}.

The pure molecular BEC at 764\,G serves as our starting point for
exploring the crossover to the degenerate Fermi gas.
Before we change the magnetic field, we first adiabatically
increase the trap power from 3.8\,mW to 35\,mW in a 200-ms
exponential ramp. The higher power provides a trap depth of
$\sim$$k_B\times\,2\,\mu$K for the atoms, which is roughly a
factor of two above the Fermi energy, and avoids spilling of the
Fermi gas produced at magnetic fields above the resonance
\cite{li2becinn}. The compression increases the peak density of
the condensate by a factor of $2.5$. All further experiments
reported here are performed in the recompressed trap with
$\omega_r/2\pi = 640\,$Hz and $\omega_z/2\pi=(600 B/{\rm kG}+
32)^{1/2}$Hz.

We measure the lifetime of the BEC in the compressed trap at 764G
to be $40$s. The peak molecular density is estimated to be
$n_{mol}=(15/8\pi)(\omega_r/\omega_z)^2N_{mol}/z_{TF}^3=1.0(5)
\times 10^{13} {\rm cm}^{-3}$. This provides an upper bound for
the binary loss coefficient of $1\times 10^{-14}$\,cm$^3$/s, and
is consistent with previous measurements in thermal molecular
gases \cite{li2thermal1,li2thermal2} together with the predicted
scattering length scaling \cite{petrovam} and the factor-of-two
suppression of binary collision loss in a condensate.

For exploring the crossover to a Fermi gas we apply slow magnetic
field ramps. To ensure their adiabaticity we performed several
test experiments. In one series of measurements we ramped up the
field from 764G to 882G and back to 764G with variable ramp speed.
This converts the molecular BEC into a strongly interacting Fermi
gas and vice versa. Therefore substantial changes are expected in
the cloud size. After the up-and-down ramp we observe an axial
oscillation of the ensemble at the quadrupolar excitation
frequency \cite{stringari, li2becinn}. This collective oscillation
is the lowest excitation mode of the system and is thus sensitive
to non-adiabaticity effects. We observe axial oscillations with
relative amplitudes of $>5\%$ for ramp speeds above $1.2$G/ms. For
ramp speeds of $0.6$G/ms and lower, the axial oscillation was no
longer visible.

\begin{figure}
\includegraphics[width=3in]{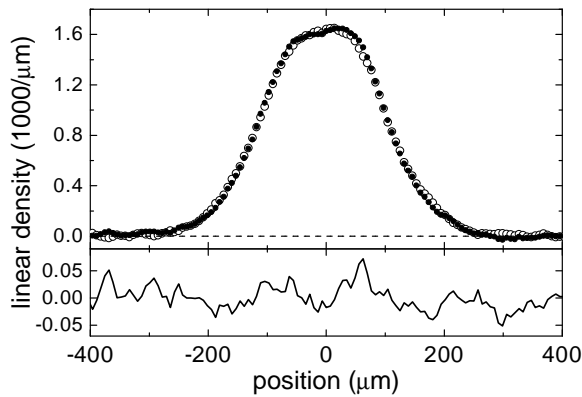}
\caption{Axial profile of a molecular BEC at 764\,G ($\bullet$)
after its conversion into a Fermi gas at 1176\,G and subsequent
back-conversion. Two 1-s magnetic field ramps are applied in this
reversible process. For reference we show the corresponding
profile observed without the magnetic field ramp ($\circ$). The
density profiles are obtained by averaging over 50 images. The
difference shown in the lower graph is consistent with the drifts
of a residual interference pattern in the images.} \label{fig2}
\end{figure}

We also checked the reversibility of the crossover process by
linearly ramping up the magnetic field from 764\,G to 1176\,G and
down again to 764\,G within 2\,s (ramp speed of $\pm$0.41\,G/ms).
In Fig.~\ref{fig2}, we compare the axial profile taken after this
ramp ($\bullet$) with the corresponding profile obtained after
2\,s at fixed magnetic field ($\circ$). The comparison does not
show any significant deviation. This highlights that the
conversion into a Fermi gas and its back-conversion into a
molecular BEC are lossless and proceed without noticeable increase
of the entropy.

To investigate the spatial profile of the trapped gas in different
regimes we start with the molecular BEC at 764\,G and change the
magnetic field in 1-s linear ramps to final values between 740\,G
and 1440\,G. Images are then taken at the final ramp field. To
characterize the size of the trapped gas, we determine the
root-mean-squared axial size $z_{\rm rms}$. This rms-size is
related to the axial radius $z_{\rm TF}$ by $z_{\rm rms} = z_{\rm
TF}/\sqrt{7}$ in the case of a pure BEC in the Thomas-Fermi limit
and by $z_{\rm rms} = z_{\rm TF}/\sqrt{8}$ in the cases of
zero-temperature non-interacting or strongly interacting Fermi
gases \cite{fit}.

\begin{figure}
\includegraphics[width=3in]{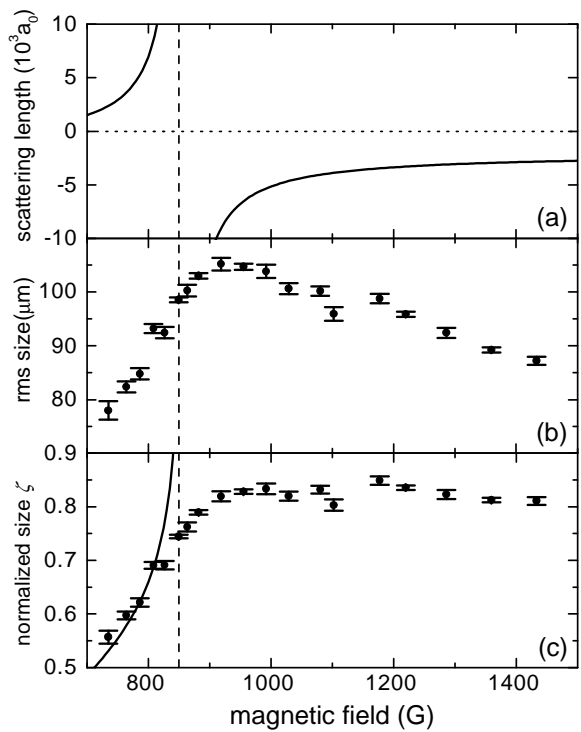}
\caption{Axial cloud size measurements across the Feshbach
resonance. In (a) the atomic scattering length $a$ is shown
according to \cite{feshbach850}; the resonance at 850\,G is marked
by the vertical dashed line. The data in (b) display the measured
rms cloud sizes. In (c), the same data are plotted after
normalization to a non-interacting Fermi gas. The solid line shows
the expectation from BEC mean-field theory with $a_{mol} =
0.6\,a$. In (b) and (c), the error bars show the statistical error
of the size measurements from typically five individual images.}
\label{fig3}
\end{figure}

Fig.\ \ref{fig3}(b) shows how the measured axial size $z_{\rm
rms}$ changes with the magnetic field. For comparison, Fig.\
\ref{fig3}(a) displays the magnetic-field dependence of the atomic
scattering length $a$. Up to 950\,G an increase in $z_{\rm rms}$
is due to the crossover from the molecular BEC to the degenerate
Fermi gas. For higher magnetic fields the axial cloud size of the
Fermi gas shrinks with $B$ as the axial magnetic confinement
increases ($\omega_z \propto \sqrt{B}$).

For the following discussions, we normalize the observed size to
the one expected for a non-interacting Fermi gas. In particular,
this removes the explicit trap dependence. In Fig.\ \ref{fig3}(c)
we show the normalized axial size $\zeta=z_{\rm rms}/z_0$, where
$z_0=(E_F/4m\omega_z^2)^{1/2}$ is the rms axial size of a
non-interacting zero-temperature Fermi gas with $N = 4\times10^5$
atoms. The Fermi energy $E_F = \hbar^2k_F^2/2m=\hbar
\bar{\omega}(3N)^{1/3}$ amounts to $k_{\rm B} \times 1.1\,\mu$K at
850\,G, and the Fermi wave number $k_F$ corresponds to a length
scale of $k_F^{-1} = 3600\,a_0$.

Below the Feshbach resonance, the observed dependence of the cloud
size agrees well with the mean-field behavior of a BEC in the
Thomas-Fermi limit. In this regime, the normalized size is given
by $\zeta=0.688 (a_{mol}/a)^{1/5} (E_F/E_b)^{1/10}$, where
$E_b=\hbar^2/ma^2$ is the molecular binding energy. Fig.\
\ref{fig3}(c) shows the corresponding curve (solid line)
calculated with $a_{mol}/a=0.6$ \cite{petrovam}. This BEC limit
provides a reasonable approximation up to $\sim$800\,G; here the
molecular gas interaction parameter is $n_{mol}a_{mol}^3 \approx
0.08$. Alternatively, the interaction strength can be expressed as
$k_Fa\approx 1.9$.

\begin{figure}
\includegraphics[width=3in]{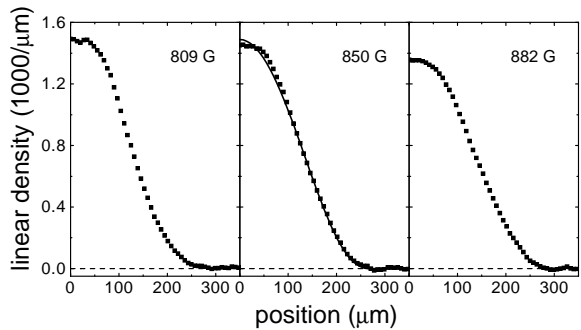}
\caption{Observed axial density profiles near the Feshbach
resonance, averaged over 50 images and symmetrized to reduce
imaging imperfections. The rms cloud sizes are 93\,$\mu$m,
99\,$\mu$m, and 103\,$\mu$m at $B= 809$G, 850G, and 882G,
respectively. For comparison, the on-resonance data at 850\,G are
shown together with a fit by the expected profile $\propto
(1-z^2/z_{\rm TF}^2)^{5/2}$. The small deviation near the top is
due to a residual interference pattern in the images.}
\label{fig4}
\end{figure}

The crossover to the Fermi gas is observed in the vicinity of the
Feshbach resonance between 800\,G and 950\,G; here $\zeta$
smoothly increases with the magnetic field until it levels off at
950\,G, where the interaction strength is characterized by $k_Fa
\approx -1.9$. Our results suggest that the crossover occurs
within the range of $-0.5\lesssim (k_Fa)^{-1}\lesssim0.5$, which
corresponds to the strongly-interacting regime. The smoothness of
the crossover is further illustrated in Fig.~\ref{fig4}. Here the
spatial profiles near the resonance show the gradually increasing
cloud size without any noticeable new features.

On resonance a universal regime is realized \cite{heiselberg,
ho03, thomasmechan}, where scattering is fully governed by
unitarity and the scattering length drops out of the description.
Here the normalized cloud size can be written as
$\zeta=(1+\beta)^{1/4}$, where $\beta$ parameterizes the
mean-field contribution to the chemical potential in terms of the
local Fermi energy \cite{thomasmechan}. At 850\,G our measured
value of $\zeta=0.75\pm0.07$ provides
$\beta=-0.68^{+0.13}_{-0.10}$. Here the total error range includes
all statistic and systematic uncertainties with the particle
number giving the dominant contribution. Note that the uncertainty
in the Feshbach resonance position is not included in the errors
\cite{feshbach850b}. Our experimental results reveal a stronger
interaction effect than previous measurements that yielded
$\beta=-0.26(7)$ at $T=0.15T_F$ \cite{thomasmechan} and
$\beta\approx -0.3$ at $T=0.6T_F$ \cite{salomon}. Our value of
$\beta$ lies within the range of the theoretical predictions for a
zero temperature Fermi gas: $-0.67$ \cite{betabaker, heiselberg},
$-0.43$ \cite{betabaker} and, in particular, $-0.56(1)$ from a
recent quantum Monte Carlo calculation \cite{carlson}.

Beyond the Feshbach resonance, in the Fermi gas regime above
$950$\,G we observe an essentially constant normalized cloud size
of $\zeta=0.83\pm0.07$. In this regime the interaction parameter
$k_Fa$ is calculated to vary between $-2$ (at 950\,G) and $-0.8$
(at 1440\,G), which allows us to estimate $\zeta$ to vary between
0.90 and 0.95 based on the interaction energy calculations in Ref.
\cite{heiselberg}. Our observed values are somewhat below this
expectation, which requires further investigation.


In summary, we have demonstrated the smooth crossover from a
molecular condensate of $^6$Li dimers to an atomic Fermi gas.
Since the conversion is adiabatic and reversible, the temperature
of the Fermi gas can be estimated from the conservation of entropy
\cite{carr}. Our high condensate fraction of $> 90\%$ suggests a
very small entropy which in the Fermi gas limit corresponds to an
extremely low temperature of $k_BT<0.04E_F$. In this scenario,
superfluidity can be expected to extend from the molecular BEC
regime into the strongly interacting Fermi gas regime above the
Feshbach resonance where $k_Fa\lesssim -1$. Our experiment thus
opens up intriguing possibilities to study atomic Cooper pairing
and superfluidity in resonant quantum gases.

We thank G.V. Shlyapnikov, W. Zwerger and S. Stringari and his
group for very useful discussions. We acknowledge support by the
Austrian Science Fund (FWF) within SFB 15 (project part 15) and by
the European Union in the frame of the Cold Molecules TMR Network
under Contract No.\ HPRN-CT-2002-00290. C.C.\ is a Lise-Meitner
research fellow of the FWF.


\begin{thebibliography}{10}
\bibitem{fit}
We fit the density profiles with the function $\rho(z) = \rho_0
(1-z^2/z_{\rm r}^2)^\alpha$, where $\rho_0$, $z_r$, and $2 \le
\alpha \le 2.5$ are free parameters. This function interpolates
between the density profile of a pure BEC in the Thomas-Fermi
limit with $\alpha=2$ and a zero-temperature non-interacting or
strongly interacting Fermi gas with $\alpha=2.5$. The rms cloud
size $z_{\rm rms}$ is obtained from $z_{\rm r}$ and $\alpha$ by
$z_{\rm rms}=z_{\rm r}/\sqrt{3+2\alpha}$.
\end{thebibliography}

\end{document}